 \documentclass[final,3p,times]{elsarticle}

\usepackage{graphicx}
 \usepackage{epsfig}

\usepackage{amssymb,hyperref}
\usepackage[nodots]{numcompress}

\newcommand{\mcdot}{\!\cdot\!}	
\newcommand{\be}{\begin{equation}}
\newcommand{\ee}{\end{equation}}
\newcommand{\bea}{\begin{eqnarray}}
\newcommand{\eea}{\end{eqnarray}}

\def\beq{\begin{equation}}
\def\eeq{\end{equation}}
\def\bea{\begin{eqnarray}}
\def\eea{\end{eqnarray}}

\begin{document}

\begin{frontmatter}



\begin{flushright}
ACFI-T13-05\\ CP3-Origins-2013-047 DNRF90\\ LA-UR-13-27904
\end{flushright}

\title{Shining LUX on Isospin-Violating Dark Matter Beyond Leading Order}


\author[LANL]{Vincenzo Cirigliano} 
\author[LANL]{Michael L. Graesser}
\author[LANL,UMASS]{Grigory Ovanesyan}
\author[LANL,CP3]{Ian M. Shoemaker}

\address[LANL]{Theoretical Division,  Los Alamos National Laboratory, 
Los Alamos, NM 87545, USA}
\address[UMASS]{Physics Department, University of Massachusetts Amherst, Amherst, MA 01003, USA}
\address[CP3]{CP$^{3}$-Origins and the Danish Institute for Advanced Study, University of Southern Denmark, Campusvej 55, DK-5230 Odense M, Denmark}

\begin{abstract}
Isospin-violating dark matter (IVDM)
has been proposed as a viable scenario to reconcile conflicting positive and null results from direct detection dark matter experiments. 
We  show that  the lowest-order dark matter-nucleus scattering rate can receive large and 
nucleus-dependent  corrections at next-to-leading order (NLO)  in the chiral expansion. 
The size of these corrections depends on the specific couplings of dark matter to quark flavors and gluons. In general the full NLO dark-matter-nucleus cross-section is not adequately described by just the zero-energy proton and neutron couplings. These statements are concretely illustrated in a scenario where the dark matter couples to quarks through scalar operators. We find the canonical IVDM scenario can reconcile the null XENON and LUX results and the recent CDMS-Si findings 
provided its couplings to second and third generation quarks either lie on a special line or are suppressed. Equally good fits with new values of the neutron-to-proton  coupling ratio are found in the presence of nonzero heavy quark couplings. CDMS-Si remains in tension with LUX and XENON10/100 but is not excluded. 
\end{abstract}


\end{frontmatter}


\section{Introduction}
\label{sect:intro}
To date, the dominant component of the matter in the Milky Way has only been detected through its gravitational interactions.  However, a number of experiments around the world are currently seeking to directly detect this Dark Matter (DM). The aim is detect the recoil energy deposited by an incident DM particle as it 
scatters on a nuclear target, producing a characteristic spectrum~\cite{Goodman:1984dc}. 

At present,  the field of DM direct detection is in an uncertain and exciting state with a number of experiments finding evidence of such a signal \cite{Bernabei:2008yi,Aalseth:2011wp}, and others seeming to exclude these same signals with null observations \cite{Akerib:2010pv,Angle:2011th,Aprile:2012nq,Akerib:2013tjd}. An apparent reconciliation however may be achieved by allowing the coupling of the DM to protons, $f_{p}$, to differ from its coupling to neutrons, $f_{n}$. While such isospin-violating Dark Matter (IVDM) has been studied by many authors~\cite{Kurylov:2003ra,Giuliani:2005my,Feng:2011vu}, it has become especially intriguing given the latest results from CDMS-Si \cite{Agnese:2013rvf}, which are na\"{i}vley at odds with the limits from XENON100 \cite{Aprile:2012nq} and LUX \cite{Akerib:2013tjd}.  For example, the authors of~\cite{Frandsen:2013cna} surveyed many different possible astrophysical and microphysical possibilities for DM and concluded that only IVDM or inelastic down-scattering significantly reduce the tension between CDMS-Si and XENON100. 
After LUX,  similar conclusions are found in Refs.~\cite{Gresham:2013mua,DelNobile:2013gba},  with ``Xenophobic"  WIMP couplings still providing a  
reconciliation of  existing results, albeit under increasing pressure. 
 
In this paper we study  the  phenomenological implications  of chiral NLO corrections to IVDM in light of the recent results by LUX~\cite{Akerib:2013tjd}.  
The chiral  corrections to WIMP-nucleus cross section have been studied in Refs.~\cite{Prezeau:2003sv,Cirigliano:2012pq}
assuming scalar WIMP-quark interactions (for axial interactions see \cite{Menendez:2012tm}). 
In contrast to the  one-nucleon-level effective field theory (EFT) developed in Ref.~\cite{Fitzpatrick:2012ix}, 
the chiral EFT  approach includes two-body effects and  is particularly well suited to connect the 
phenomenological  bounds on WIMP-nucleus cross sections to  the WIMP-quark short-distance  couplings,   
controlling other aspects of WIMP phenomenology (indirect detection, production at colliders). 
In  \cite{Cirigliano:2012pq}  it was found that for generic isospin-conserving WIMP-quark  couplings the magnitude of the NLO effects 
is of the size expected from  chiral power counting $\sim m_{\pi}/(1\rm{GeV})\sim 10\%$. 
However in the case of isospin-violating couplings at the canonical  IVDM point $r \equiv f_n/f_p  \simeq - 0.7$, 
where the signal for  Xe  is  suppressed at LO by several orders of magnitude,   it was found that the chiral corrections wash out the LO cancelation generically, 
and move the   ``Xenophobic"   point to other regions in the parameter space of WIMP-quark couplings. 
In this letter  we explore in detail these points. 

The remainder of this paper is organized as follows. In Sec.~\ref{sect:setup} we review and update our results on scalar-mediated DM-quark interactions, including now the momentum dependence in the two-body amplitude. In Sec.~\ref{sect:pheno1} we study the degradation in sensitivity experienced by a Xenon target at NLO and compare 
the effect of chiral corrections for Xenon, Silicon and Germanium targets. 
In Sec.~\ref{sect:deg}  we  discuss  parameter degeneracies and the role of hadronic and nuclear uncertainties. 
Then in Sec.~\ref{sect:pheno2} we compute the best-fit and excluded regions from the CDMS-Si,  XENON, and LUX experiments respectively. There we find that the well-known $r=-0.7$ only maintains a partial compatibility {\it provided} either that the strange and heavy quark couplings in the effective low-energy theory are sufficiently suppressed, or that these couplings lie on a line corresponding to an approximate degeneracy in the total recoil rate. In addition, we also find new regions of partial compatibility for which  $f_{n}/f_{p}$ is significantly different from $-0.7$. Finally in Sec.~\ref{sect:conclusions} we discuss the implications of these findings for future DM data, including direct detection and collider searches.

\section{Setup}
\label{sect:setup}

Below the scale of the heavy quarks, the scalar  interaction of  WIMPs (denoted by $X$) with light quarks is given by the 
effective Lagrangian~\cite{Cirigliano:2012pq}
\be
{\cal L}_{\rm eff} = \sum_{q = u,d,s} \   \frac{\lambda_q}{v  \Lambda^2}  \ \overline{X} X  \  m_q \overline{q} q   + 
 \frac{\lambda_\theta}{v  \Lambda^2}  \ \overline{X} X  \  \theta^\mu_\mu~, 
\label{eq:Leff}
\ee
where $\Lambda$ is a generic new physics scale, $v= (\sqrt{2} G_F)^{-1/2}$ is the electroweak scale   and  $\theta^\mu_\mu$ is the trace of the energy-momentum tensor. 
The effect  of WIMP couplings to heavy quarks  is encoded in the coefficient $\lambda_\theta=(2/27) \sum_Q \tilde{\lambda}_Q-(8/9) \tilde{\lambda}_G$, and also in the couplings of the light quarks through the relation $\lambda_q=\tilde{\lambda}_q - \lambda_\theta$. Here $\tilde{\lambda}_{q,Q}$ and $\tilde{\lambda}_G$ are the short-distance couplings of dark matter to light quark, heavy quarks, and the gluon field strength. 

At leading order (LO)  in chiral EFT,  the four quark-level couplings $\lambda_{u,d,s,\theta}$ collapse  
into two independent combinations, i.e. the zero momentum transfer matrix elements of ${\cal L}_{\rm eff}$ in the proton and neutron, $f_{p,n}$,  
\be
\label{eq:fpn}
f_{p,n}  =      \frac{1}{v \Lambda^2}  \,  \Big[        \sigma_{\pi N} \left(   \lambda_+  \pm  \lambda_-  \xi  \right)   + \lambda_s \sigma_s  + \lambda_\theta   \, m_p   \Big]~, 
\qquad \qquad 
\lambda_{\pm} =  (\lambda_u m_u  \pm \lambda_d m_d)/(m_u + m_d)~,
\ee
where   $\sigma_{\pi N} =( (m_u + m_d)/2  )  \langle p | \bar{u} u + \bar{d} d |p \rangle$, 
$\xi  =   \langle p | \bar{u} u -  \bar{d} d |p \rangle/  \langle p | \bar{u} u + \bar{d} d |p \rangle$,  
$\sigma_s = \langle p |  m_s \bar{s} s | p \rangle$, 
and the upper (lower)  sign refers to $p$ ($n$)~\footnote{For the 
nucleon sigma-terms we use the lattice QCD ranges 
$\sigma_{\pi N}=  45 (15)$~MeV,    $\sigma_{s}=  45 (25)$~MeV (from the review~\cite{Kronfeld:2012uk}).  
$\xi$ can be related to $y \equiv 2 \langle p |\bar{s} s | p\rangle /  \langle p | \bar{u} u + \bar{d} d |p \rangle$ 
through an analysis of  baryon masses in the  $SU(3)$ limit~\cite{Cheng:1988im}, leading to 
$\xi = (1 - y) \, 0.197  =  0.18(1)$.}.
These relations are valid up to small isospin-breaking effects of order $(m_u-m_d)/\Lambda_{QCD}$. 
Working to LO in chiral EFT,  it is convenient to trade $f_{n,p}$ for 
$\sigma_p \equiv  2\,k_X\,\mu^2f_p^2/\pi$  and $r \equiv f_n/f_p$
($\mu$ is the WIMP-proton reduced mass and   $k_X=1/2  (2)$ for Dirac (Majorana)  fermions).   
To LO the  WIMP-nucleus differential rate is then given by: 
\be
\frac{dR}{dE_R}^{\rm{LO}}=\frac{\sigma_p\, \rho_{0}}{2\mu^2m_X} \ \Big| \big(Z +(A-Z)r\big) F(E_R) \Big|^2
\times     \,\eta\left(E_R,m_X,m_A\right) ~,
\label{eq:rate1}
\ee
where $m_X$ and $m_A$ are the WIMP and target nucleus masses, $F (E_R)$ is the one-body nuclear form factor, $\rho_{0}$ is the local DM mass density, and $\eta (E_R, m_X, m_A)$ is the flux factor involving 
an integral over the  local WIMP velocity distribution~\cite{Smith:1988kw,Jungman:1995df,Savage:2006qr,McCabe:2010zh}. 
This is the familiar result used in phenomenological applications. 
Note that any value of $\sigma_p$ and  $r$ can be obtained by an appropriate choice of the quark couplings 
$\lambda_i/\Lambda^2$.  However, in the limit $\xi \to 0$ only $r=1$ is possible for all choices of $\lambda_i$, as seen from Eq.~(\ref{eq:fpn}).

As discussed in Ref.~\cite{Cirigliano:2012pq}, at  next-to-leading order (NLO)  one 
needs all  {\it four}  $\lambda_{u,d,s,\theta}$   parameters to describe the scattering rate. 
The  $\lambda_{u,d,s,\theta}$ couplings appear in the  recoil energy dependence  of neutron and proton matrix elements, 
as well as a new two-body contribution to the amplitude ($A_2 (E_R)$).    In order to make contact with the existing phenomenology we choose as independent parameters 
 the ``standard"  quantities $\sigma_p$ and $r$, as well as  the  
rescaled strange and gluonic (heavy quark) couplings $\overline{\lambda}_{s,\theta}   \equiv \lambda_{s,\theta}/\lambda_{u}$.    
With this choice, the NLO   WIMP-nucleus differential rate reads 
\be
\frac{dR}{dE_R}^{\rm{NLO}}=\frac{\sigma_p\, \rho_{0}}{2\mu^2m_X} \ \bigg| \Big[ Z  \, \big(1 +  s_p  E_R \big)    +  \big( A-Z \big)  \,    \big(r + s_n E_R\big)    \Big] \,   F(E_R)   +   A_2 (E_R)   \bigg|^2
\times     \,\eta\left(E_R,m_X,m_A\right) ~,
\label{eq:rate1nlo}
\ee
where
\bea
s_p &=&  f_u  [r, \overline{\lambda}_{s,\theta}]  \  \big( s_p^u   + f_d   [r, \overline{\lambda}_{s,\theta}]   \, s_p^d +  s_p^s \, \overline{\lambda}_s \big) \cdot  A          \\
s_n &=&  f_u    [r, \overline{\lambda}_{s,\theta}]   \  \big(s_n^u   + f_d    [r, \overline{\lambda}_{s,\theta}]   \, s_n^d +  s_n^s \, \overline{\lambda}_s  \big)   \cdot  A   \\ 
A_2 (E_R) &=&    f_u   [r, \overline{\lambda}_{s,\theta}]    \   \Big[   \big(t_u   + f_d   [r, \overline{\lambda}_{s,\theta}]    \, t_d  \big)     \, F_{\pi \pi}(E_R)     +     t_s \, \overline{\lambda}_s  \, F_{\eta \eta}(E_R)   \Big]  \cdot A ~,
\eea
and the common factor of $A$ arises in $s_{p,n}$ from $q^2 = 2 m_A E \propto A$.
The quantities $f_{u,d}   [r, \overline{\lambda}_{s,\theta}] $ arise in the change of variables from   $\lambda_{u,d} $  to  $f_p$ and $r$. 
$f_d$ is the ratio $\lambda_d/\lambda_u$ expressed in terms of the independent variables $r, \overline{\lambda}_{s,\theta}$.  
Similarly, $f_u$ represents the ratio $\lambda_u/(v\Lambda^2 f_p)$ expressed in terms of $r, \overline{\lambda}_{s,\theta}$. 
The explicit form of $f_{u,d}$ depends not only on $r,  \overline{\lambda}_{s,\theta}$  but also on  the hadronic matrix elements appearing in  (\ref{eq:fpn}): 
\bea
\label{eq:fu}
f_u  &=& \frac{ 1 + \xi   - r (1 - \xi)}{2 \,  \xi \, \left[ \delta f  +  \frac{2 m_u}{m_u+m_d}  \, \sigma_{\pi N} \right]}~,   \qquad \qquad \qquad 
\delta f =  \overline{\lambda}_s \sigma_s  + \overline{\lambda}_\theta m_p~,    \\
f_u f_d &=&  \frac{ (r- 1) \delta f   -  \frac{m_u}{m_u + m_d}     \sigma_{\pi N}  \left[ 1 - \xi - r (1 + \xi)\right] }{
2 \,  \xi \, \left[ \delta f  +  \frac{2 m_u}{m_u+m_d}  \, \sigma_{\pi N} \right]    \frac{m_d}{m_u + m_d}     \sigma_{\pi N}    }~.
\label{eq:fd}
\eea
Note that there is an apparent singularity in the above expressions when the denominators  vanish.   
This corresponds  to the limit $f_p \to 0$.  In that case the fractional correction diverges,  but that's simply because we are factoring out $f_p$. 
The  coefficients appearing in $s_{p,n}$ and $A_2(E_R)$ 
are known from the NLO EFT analysis of Ref.~\cite{Cirigliano:2012pq} and are 
reported in Table~\ref{tab:inputs1}~\footnote{Note that the numerical values of $t_u, t_d, t_s$ depend on the nuclear matrix 
elements $N_{\pi\pi}(0)$ and $N_{\eta\eta}(0)$~\cite{Cirigliano:2012pq}.   In Ref.~\cite{Cirigliano:2012pq} these were computed within the shell model, using 
an unconventional cut on relative nucleon distance of $d_c=2^{1/2} \mcdot \,0.5\,\rm{fm}$, leading to  
$N_{\pi\pi}(0)=-0.91 A$ and $N_{\eta\eta}(0)=0.0061 A$.
Here we use the more  conventional cut  $d_c=0.5\,\rm{fm}$, resulting in $N_{\pi\pi}(0)=-1.19\,A$ and $N_{\eta\eta}(0)=0.0048\,A$.}.
Extending the work in Ref.~\cite{Cirigliano:2012pq}, within the shell model we include here the    recoil energy   dependence of the two-body amplitude:
\begin{eqnarray}
F_{\pi \pi} (E_R) &=&F_{\rm{exp}}  \left[ \left(1.20-1.83\,A^{-1/3}+4.60\,A^{-2/3}\right)\mcdot|{\bf{q}}|\right]~, \qquad \qquad  |\bf{q}|  =  \sqrt{2   m_A  E_R}   \\
F_{\eta \eta} (E_R)&=&F_{\rm{Bessel}}\left[ \left(0.74+1.04\,A^{-1/3}-1.93\,A^{-2/3}\right)\mcdot|{\bf{q}}|\right]~. 
\end{eqnarray}
In the above expressions we have 
$ F_{\rm{exp}}(q) = \exp\left(-q^{2}R_{0}^{2}/6\right)$ with $R_{0} \equiv \left[0.3 + 0.91\left(m_{A}/{\rm GeV} \right)^{1/3}\right]~{\rm fm}$,
and 
$F_{\rm{Bessel}}(q) = 3\left(   (\sin(qr_{n}) - qr_{n} \cos(qr_{n}))/  (qr_{n})^{3} \right)\times e^{-(qs)^{2}/2}$ with $r_{n} \equiv A^{1/3}~{\rm fm}, ~~s = 1~{\rm fm}$. 
This form was found by computing for  closed shells ($A=4, 16, 40, 80, 140$)  and then fitting the result with one-body form factors  
with $A$-dependent rescaling of the argument.

\begin{table}[t!]\begin{center}
\vspace{0.2cm}
\begin{tabular}{|c|c|c|c|c|c|c|c|}
\hline  
$s_p^u$  & 
$s_p^d$ &  
$s_n^u$ &  
$s_n^d$ &
$s_{p,n}^s$ & 
$t_u$ & 
$t_d$ &
$t_s$ \\
\hline
\hline 
 - 0.116      &   -0.192  &  -0.096 & -0.232  &  -0.472 &  -0.63~MeV   &  -1.27~ MeV  &  0.070~MeV \\
\hline
\end{tabular}
\caption{\it \small 
Numerical values of the coefficients entering the NLO  amplitude.
The uncertainty in the combination of low-energy constants $F/(F+D) \in [0.3,0.5]$ affects $s_N^{u,d}$ at the 5\% level and $s_N^s$ at the 20\% level~\cite{Cirigliano:2012pq}.
The dimensionful  two-body coefficients $t_{u,d,s}$  have been estimated through a nuclear shell model calculation 
in  Ref.~\cite{Cirigliano:2012pq}, and are in principle subject to larger uncertainties.}
\label{tab:inputs1}
\end{center}
\vspace{-0.5cm}
\end{table}

\section{Degradation factors beyond leading order}
\label{sect:pheno1}

Scalar-mediated interactions induce coherent WIMP-nucleus scattering, 
which for $f_p \sim f_n$ implies the well-known overall factor of $A^2$ in the cross-section. 
In general,  for $f_p \neq f_n$ interference effects can suppress the cross-section relative to the 
case $f_p=f_n$,  and a useful measure of this suppression is provided by the 
so-called degradation factor~\cite{Feng:2013vod,Feng:2013vaa,Kopp:2011yr,Schwetz:2011xm}. 
The original references worked to LO in ChPT and their definition 
can be cast in terms of the integrated rates $\bar{R}$ as 
\be
D^{LO}  (r)   =  \frac{\overline{R}^{LO} \left( r, \sigma_p \right)}{ \overline{R}^{LO} \left( 1, \sigma_p \right)}~, 
\qquad \qquad 
\overline{R}   \equiv  \int_{E_R^{\rm min}}^{E_R^{\rm max}}  \, dE_R  \ \frac{d R}{dE_R}~,  
\ee
with experiment-dependent integration limits $E_R^{\rm min / max}$. 
Note that for a given isotope  $D^{LO} \propto [Z  + (A-Z) r]^2 $   and 
one can use either the integrated or the differential rate, as the energy-dependence cancels in the ratio. 
This is not true anymore to NLO, so  we generalize the definition of degradation factor as follows 
\be
D^{NLO}  (r, \overline{\lambda}_s, \overline{\lambda}_\theta)   =  \frac{\overline{R}^{NLO} \left( r, \sigma_p, \overline{\lambda}_s, \overline{\lambda}_\theta \right)}{ \overline{R}^{LO} 
\left( 1, \sigma_p\right)} ~, 
\label{DNLO}
\ee 
and note that while the dependence on $\sigma_p$ drops in the ratio,  
$D^{NLO}$ depends not only on $r$, but also on $\overline{\lambda}_{s, \theta}$.  

Inspection of Eqs.~(\ref{eq:rate1nlo}) through (\ref{eq:fd})  shows that  $D^{NLO}$ is still a quadratic form in $r$.   
However,  as illustrated below, for a given target   the location of the minimum and the 
value at the minimum are affected in a non-trivial way by the chiral corrections.

\begin{figure}[!t]
\begin{center}
\includegraphics[width=0.32\textwidth]{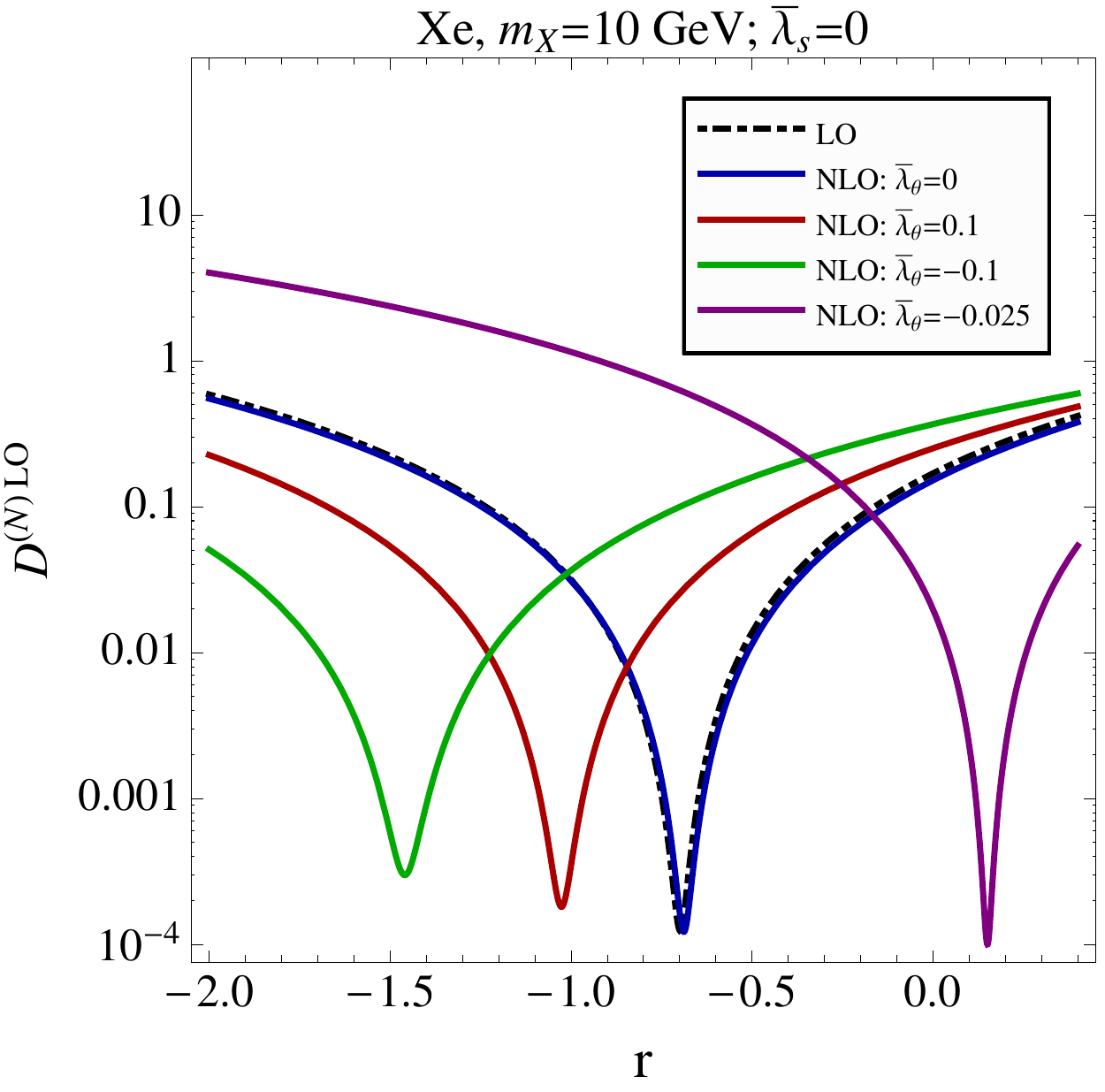}
~~~\includegraphics[width=0.31\textwidth]{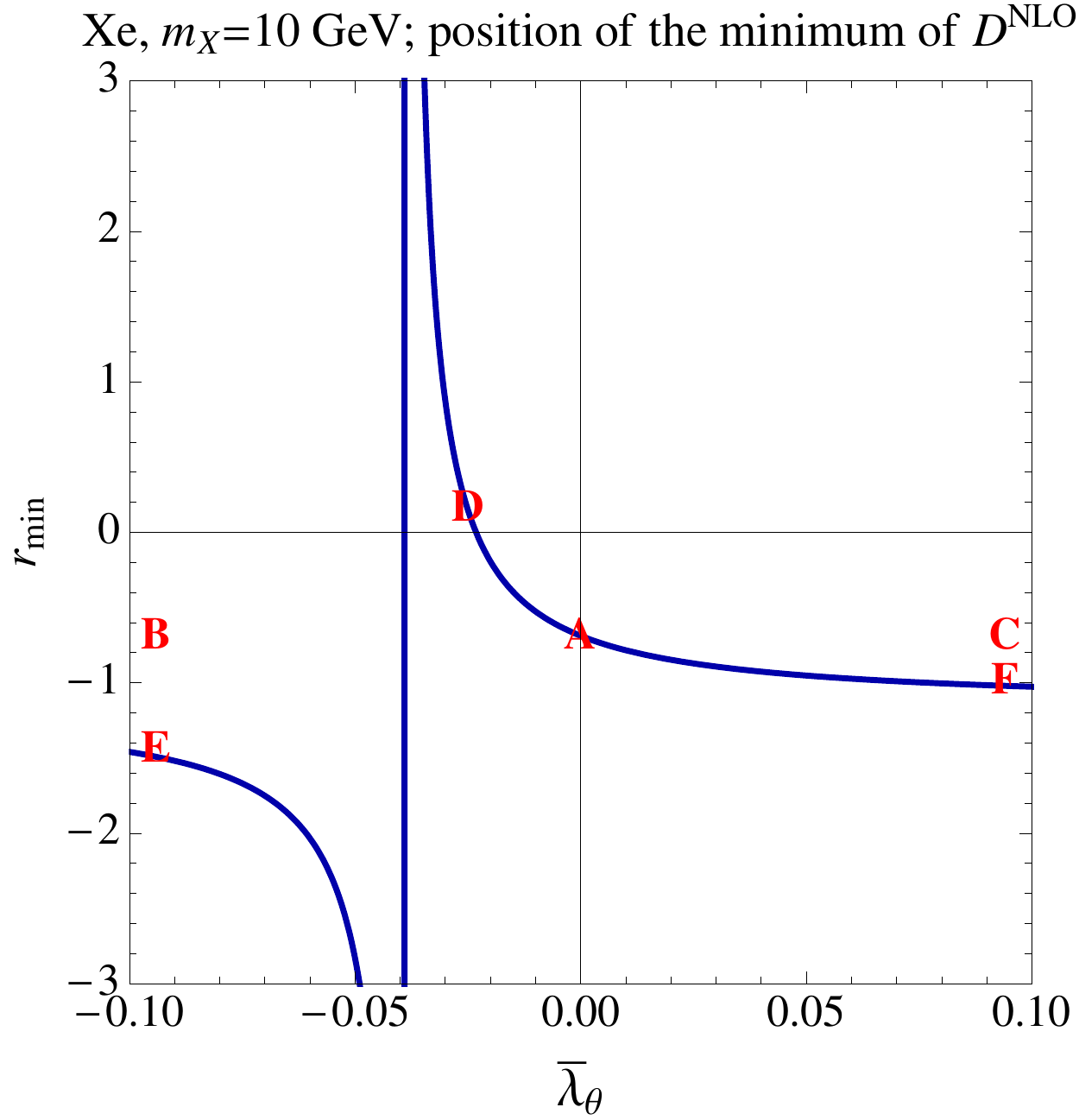}
~~~\includegraphics[width=0.325\textwidth]{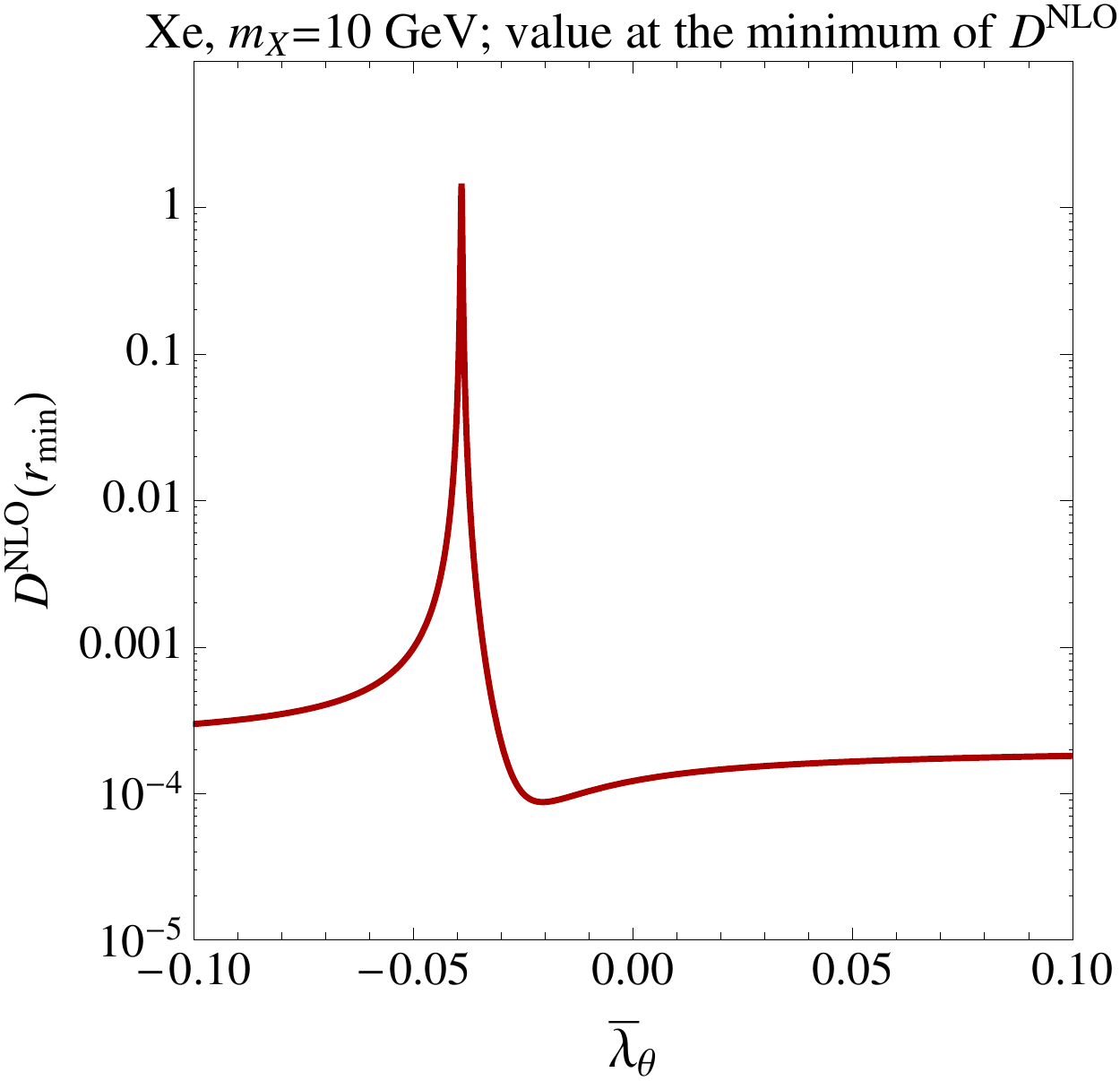}
\vspace{-.3cm}
\caption{\label{fig:pheno1} 
{\it Left panel}: Xenon degradation factors. 
Solid lines represent   $D^{NLO} (r,\overline{\lambda}_s,\overline{\lambda}_\theta)$
(Eq.~(\ref{DNLO}))  with $\lambda_{\theta}= \lambda_{s}=0$ (blue line), $\overline{\lambda_{\theta}}=0.1$ (red line),  $\overline{\lambda_{\theta}}=-0.1$ (green line), 
and $\overline{\lambda_{\theta}}=-0.025$ (purple  line). 
The dashed blue line represents $D^{LO}(r)$. 
$D^{LO}(r)$ and $D^{NLO} (r,0,0)$ 
are nearly degenerate, as explained in the text. Note that for other values of $\lambda_s$ 
and $\lambda_{\theta}$ the degradation factor at NLO has a sizable shift. 
{\it Middle panel}: Dependence of the position of the minimum of $D^{NLO}$, denoted by  $r_{min}$,  on $\overline{\lambda}_{\theta}$, with $\lambda_s=0$.  Benchmarks discussed further in the text are also shown. {\it Right panel}:  Dependence of  the value of $D^{NLO}(r_{min})$ on $\overline{\lambda}_{\theta}$ with $\lambda_s=0$.  Note that at  
$r_{\rm min}$  the values of the degradation factor are nearly independent of $\lambda_{\theta}$. 
}
 \vspace{-0.5cm}
\end{center}
\end{figure}

In Fig.~\ref{fig:pheno1} we illustrate the impact of chiral corrections on the degradation factor, using as a benchmark  the Xenon target (summing over isotopes).    
In the left panel we show both $D^{LO}$ (dashed line) and $D^{NLO}$ versus $r$ for
$\overline{\lambda}_{s} = 0$  and  $ \overline{\lambda}_\theta =0,\pm 0.1$. 
A few salient features emerge: 
first, in the absence of 2$^{nd}$ and 3$^{rd}$ generation couplings (in the low-energy theory)
the NLO corrections are \%-level and  do not significantly affect the degradation factors~\footnote{This can be understood as follows: 
in the region $r \sim -1$ one finds  $f_d \sim -m_u/m_d \sim -1/2$, 
which combined with the numerical values in Table~\ref{tab:inputs1} simultaneously suppresses  both the slopes $s_{p,n}$ and $A_2$, 
i.e. the entire NLO corrections. 
In the region  $r  \neq - 1$ the suppression comes from the overall factor $f_u$, that gets 
suppressed by a factor of $\xi \sim 0.18$ compared to its value at $r \sim -1$.}.
However, as one ``turns on" the WIMP coupling to strange and $\theta^\mu_\mu$,  even at a level of 10\%  of the light quark couplings,  the results change dramatically, 
with an $O(1)$ shift in the value  of $r$ for which the degradation factor has a dip  
(compared to the well-known LO case $r \simeq -0.7$). 
The bulk of the shift is caused by the two-body correction $A_2$ in Eq.~(\ref{eq:rate1nlo}), 
as one can  verify using  Eqs.~(\ref{eq:rate1nlo}) through (\ref{eq:fd})  and typical recoil energies of $O(10)$~keV. 
That  the NLO corrections  depend on $\overline{\lambda}_\theta$ may at first seem strange, since they do not have any such explicit dependence.  Such a dependence is induced through our choice of independent parameters (namely  $\overline{\lambda}_d \equiv \lambda_d/\lambda_u$  depends not only on $r$, but also on $\overline{\lambda}_s$ and $\overline{\lambda}_\theta$).
 
Varying $\overline{\lambda}_s$  while keeping $\overline{\lambda}_\theta=0$  produces similar results.   
In fact, neglecting the small slope corrections, the effect of  $\overline{\lambda}_{s,\theta}$   
is degenerate, as they appear in the linear combination $\delta f =  \sigma_s \overline{\lambda}_s   + m_p   \overline{\lambda}_\theta$. 
Finally, we note that   sizable shifts in the minimum location arises when varying the nucleon sigma term $\sigma_{\pi N}$.  
We will discuss in greater detail these degeneracies and hadronic uncertainties in Sec.~\ref{sect:deg}. 

Given the sensitivity to the strange quark and $\theta^{\mu}_ \mu$ couplings demonstrated above, it is interesting to track the 
location and depth of the 
``dip"  in  the degradation factor as a function of  $\overline{\lambda}_{s,\theta}$.    
We illustrate this variation in the middle and right panels of  Fig.~\ref{fig:pheno1}.  
One can see that at NLO the dip can occur at virtually any value of $r$ (even positive values!)  provided we adjust $\overline{\lambda}_\theta$ accordingly. In the middle panel we indicate six benchmark scenarios $A,B,C,D,E,F$ in the $r,\overline{\lambda}_{\theta}$ plane. 
We chose them in such a way that  $A$ is the canonical IVDM scenario ($r=-0.7$, $\overline{\lambda}_s=\overline{\lambda}_\theta=0$)
while  $B$ and $C$ are perturbations around it. 
The last three benchmarks  correspond to plausible reconciliation of  XENON/LUX  with CDMS-Si. Values of $r$ are $r=-0.7$ for $A,B$ and $C$, $r=+0.15$ for $D$, $r=-1.45$ for $E$, and $r=-1$ for $F$. We will come back to these benchmark scenarios in the  Sec~\ref{sect:pheno2}. 
The right panel of Fig.~\ref{fig:pheno1} shows that for most  values of $r_{\rm min}$   the degradation is close to $10^{-4}$, 
suggesting that  indeed there is a manifold of ``Xenophobic"  couplings in which XENON/LUX  exclusion regions  might be consistent with 
signals claimed in  experiments using Ge or Si targets. 

\begin{figure}[!t]
\begin{center}
\includegraphics[width=.4\textwidth]{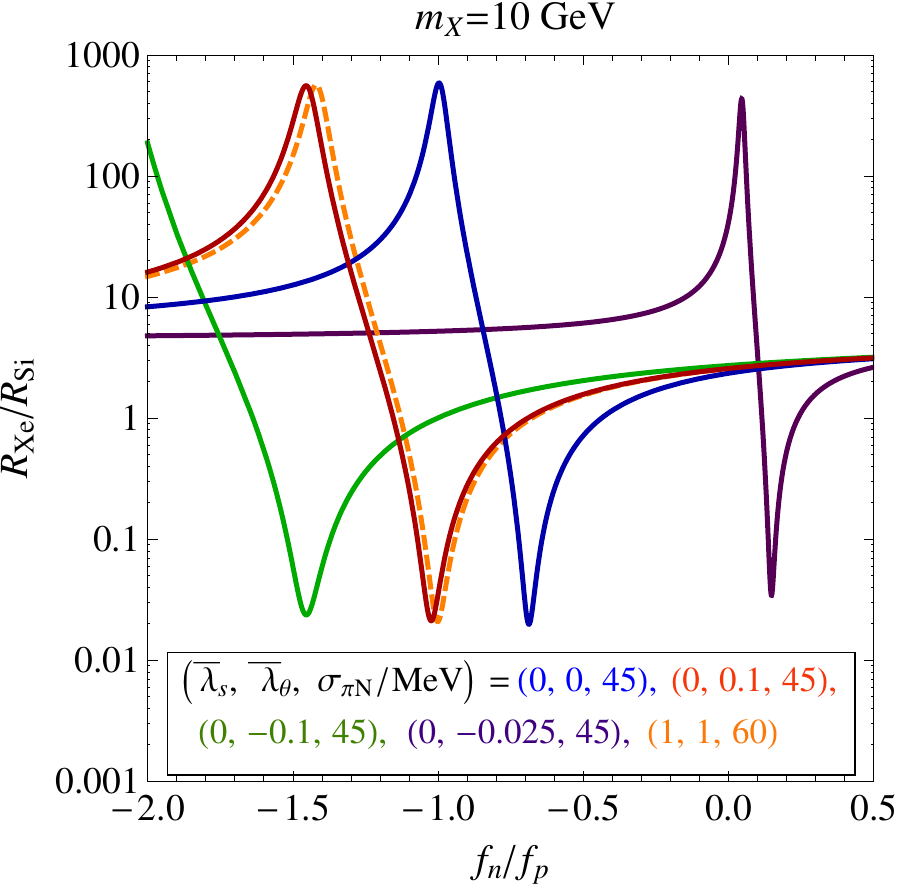}
~~~~~~~~~~~\includegraphics[width=0.4\textwidth]{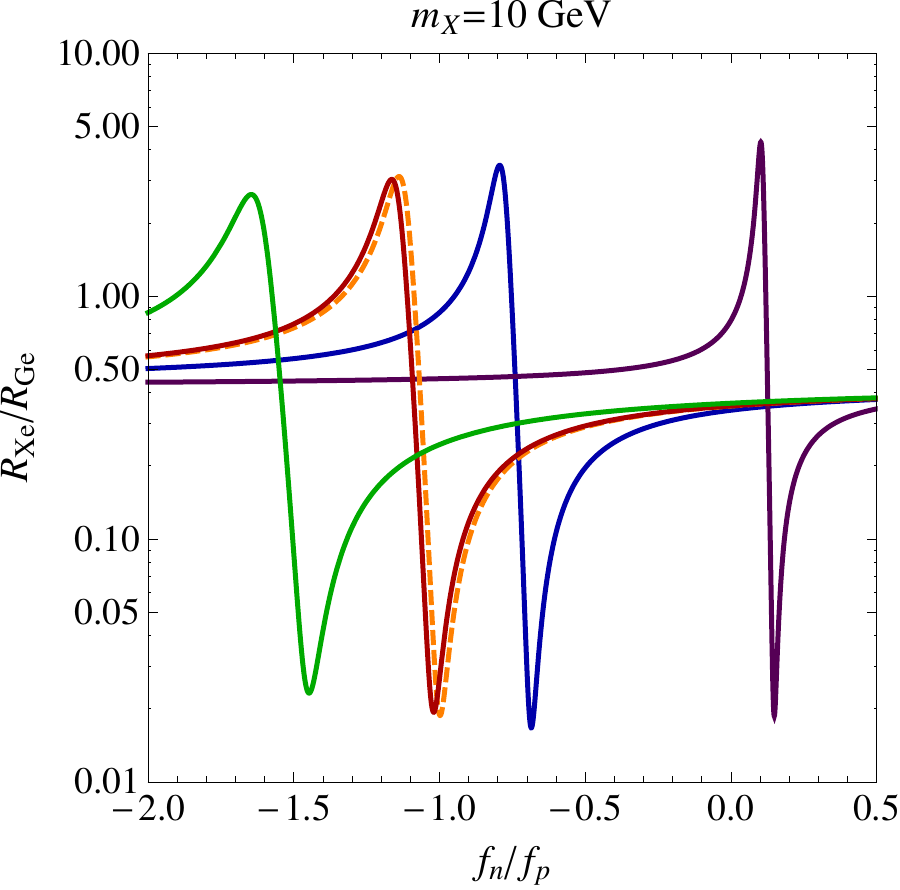}
\vspace{-.3cm}
\caption{\label{fig:pheno2} 
Double ratio of  total rates   $\bar{R}^{NLO} (Xe)/ \bar{R}^{NLO} (Si)$  (left panel) and 
$\bar{R}^{NLO} (Xe)/ \bar{R}^{NLO} (Ge)$  (right panel) versus $r$,    
for $\overline{\lambda}_{s} = 0$, $\sigma_{\pi N}=45$ MeV,  and  $ \overline{\lambda}_\theta =0,-0.025,\pm 0.1$. Also shown is 
the double ratio for $\sigma_{\pi N}=60$ MeV and $\overline{\lambda}_{s,\theta}=1 $ (dashed red). Note the degeneracy of this curve with the one for $\sigma_{\pi N}=45$ MeV and  
$\overline{\lambda}_\theta = 0.1$. Such degeneracies are described in further detail in Sec. \ref{sect:deg}.}
 \vspace{-0.5cm}
\end{center}
\end{figure}

To make the latter point plausible, however, one needs to check that 
Xenon degradation at NLO is not accompanied by excessive degradation in other targets. 
To this end,  we  plot in Fig.~\ref{fig:pheno2}  the ratio of  integrated rates    $\bar{R}^{NLO} (Xe)/ \bar{R}^{NLO} (Si)$  (left panel) and 
$\bar{R}^{NLO} (Xe)/ \bar{R}^{NLO} (Ge)$  (right panel) versus $r$,    
for $\overline{\lambda}_{s} = 0$  and  $ \overline{\lambda}_\theta =0,-0.025,\pm 0.1$. 
In obtaining these plots we use experiment-specific energy windows, corresponding to LUX $([3,30]$ keV),  CDMS-Si $([7,100]$ keV), 
and CDMS-Ge $([2,100]$ keV).  While the fine details 
might change when using different energy thresholds,  
Fig.~\ref{fig:pheno2}  strongly supports the existence of a manifold of ``Xenophobic"  couplings consistent with current data.

Based on these results, we expect two qualitative changes in the phenomenology of IVDM: 
(1) turning on sizable nonzero strange quark and/or $\theta^\mu_\mu$ couplings with $r = -0.7$  generically worsens the compatibility of CDMS-Si and XENON/LUX  
such that these couplings are excluded,  and (2)  new regions of compatibility arise in which $r \neq  -0.7$.
In other words: the notion of ``Xenophobic" couplings extends beyond the $r = -0.7$ point.  
Compatibility regions might be obtained for any value of $r$,  by turning on specific couplings of the WIMP to heavy quarks or gluons. 
This makes the IVDM scenario  far richer, but of course more model-dependent. 
In Sec.~\ref{sect:pheno2}  we investigate these possibilities further with a more detailed examination of the CDMS-Si,  CDMS-Ge, XENON10/100, and LUX data.

\section{Parameter degeneracies,  hadronic uncertainties, and higher order corrections} 
\label{sect:deg} 

In this section we provide an analytic description of  how the dominant NLO chiral corrections affect  
direct-detection phenomenology.  This explains the observed degeneracies in the $\overline{\lambda}_{s,\theta}$ parameter space 
and allows us to assess the impact of hadronic and nuclear uncertainties, and higher order corrections.
While in the numerical studies we use the full NLO corrections, in this Section we obtain an approximate analytic solution by keeping only the dominant NLO effects. This means we:  
neglect (i) all slope terms compared to the two-body corrections ($s_N^q  E_R  \ll t_{u,d,s}$);  (ii) ignore the
strange contribution to 2-body amplitude ($t_s \ll t_{u,d}$);  (iii) and drop terms of $O(\xi)$  compared to terms of $O(1)$. 
With these assumptions we find that the NLO corrections  are controlled by the quantity $\Delta$: 
\bea
\label{eq:rate2nlo}
\frac{dR}{dE_R}^{\rm{NLO}} & \propto  &  
\bigg|   \Big[  Z  F(E_R)  +  A \, \Delta  \,  \ F_{\pi \pi} (E_R) \Big]     +   r  \,  \Big[ \left(A  - Z \right)  F (E_R)  -  A \,  \Delta  \,   \  F_{\pi \pi} (E_R) \Big] \bigg|^2  
\\
\Delta &=& 
  \frac{1}{2 \,  \xi \, \left[ \frac{\delta f}{\sigma_{\pi N}}   +  \frac{2 m_u}{m_u+m_d}  \right]}
 \cdot 
  \left[   \frac{t_u}{\sigma_{\pi N}}
  -  \frac{t_d}{\sigma_{\pi N}}
  \frac{m_u + m_d}{m_d}     \left(\frac{\delta f}{\sigma_{\pi N}}   + \frac{m_u}{m_u + m_d} \right)  
  \right]~. 
\eea
Setting $F(E_R) =  F_{\pi \pi} (E_R)  = 1$  (which is a good approximation for light WIMPs) we obtain 
for the location of the minimum  
\be
r_{\rm min} =  - \frac{\bar{Z}}{1 - \bar{Z}} \cdot \frac{1 + \frac{\Delta}{\bar{Z}}}{1 - \frac{\Delta}{1 - \bar{Z}}}  \qquad \qquad \qquad  \bar{Z} = Z/A~,
 \label{eq:rmin}
\ee
where the first factor is the LO  result and the second factor represents the NLO  shift. 
After appropriate averaging over multiple isotopes,  the above expressions explain quite  accurately 
the  corrections  we observe in our parameter scan.

\begin{figure}[!t]
\begin{center}
\includegraphics[width=0.4\textwidth]{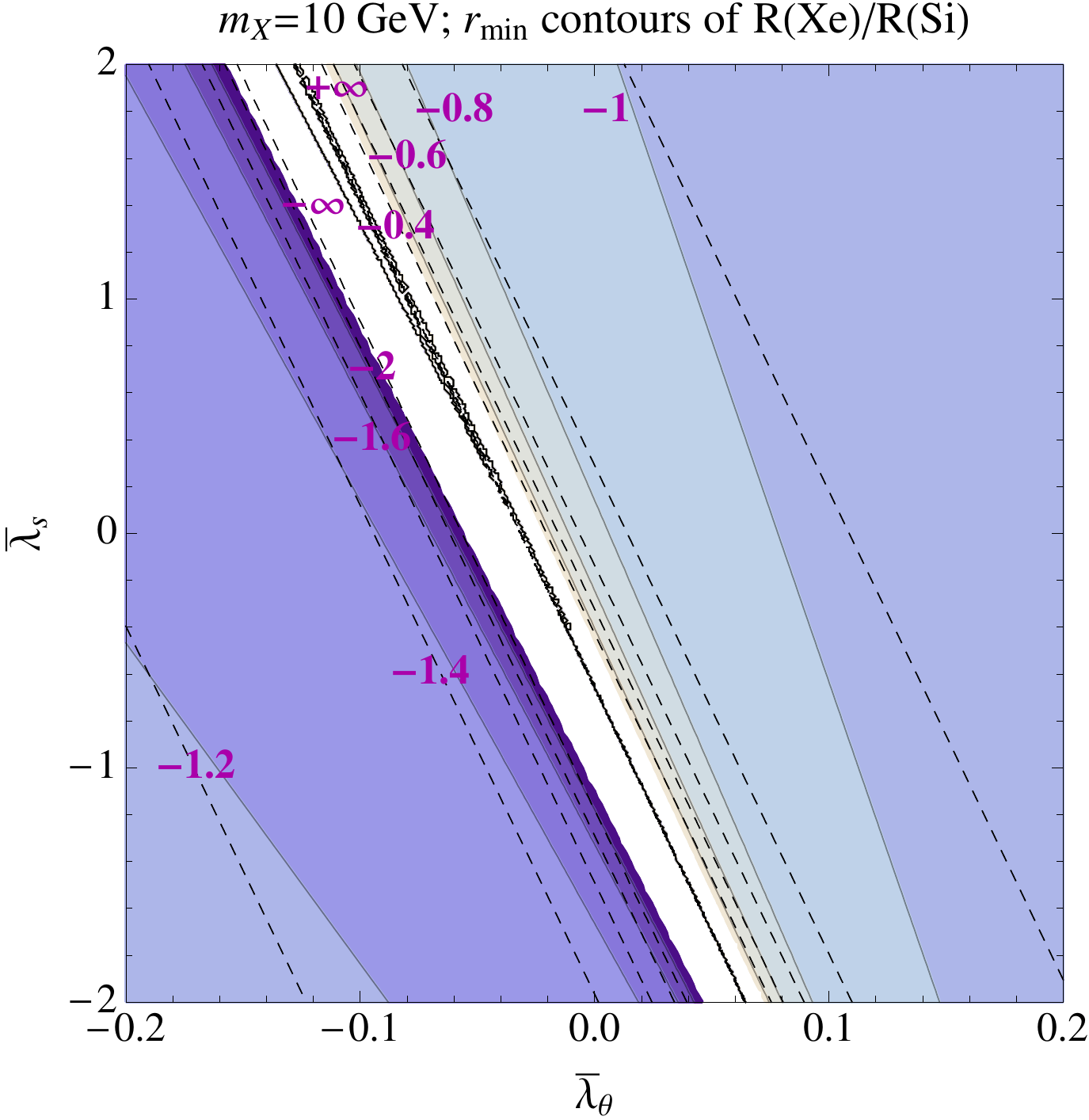}
~~~~~~~~~~~~~ 
\includegraphics[width=0.4\textwidth]{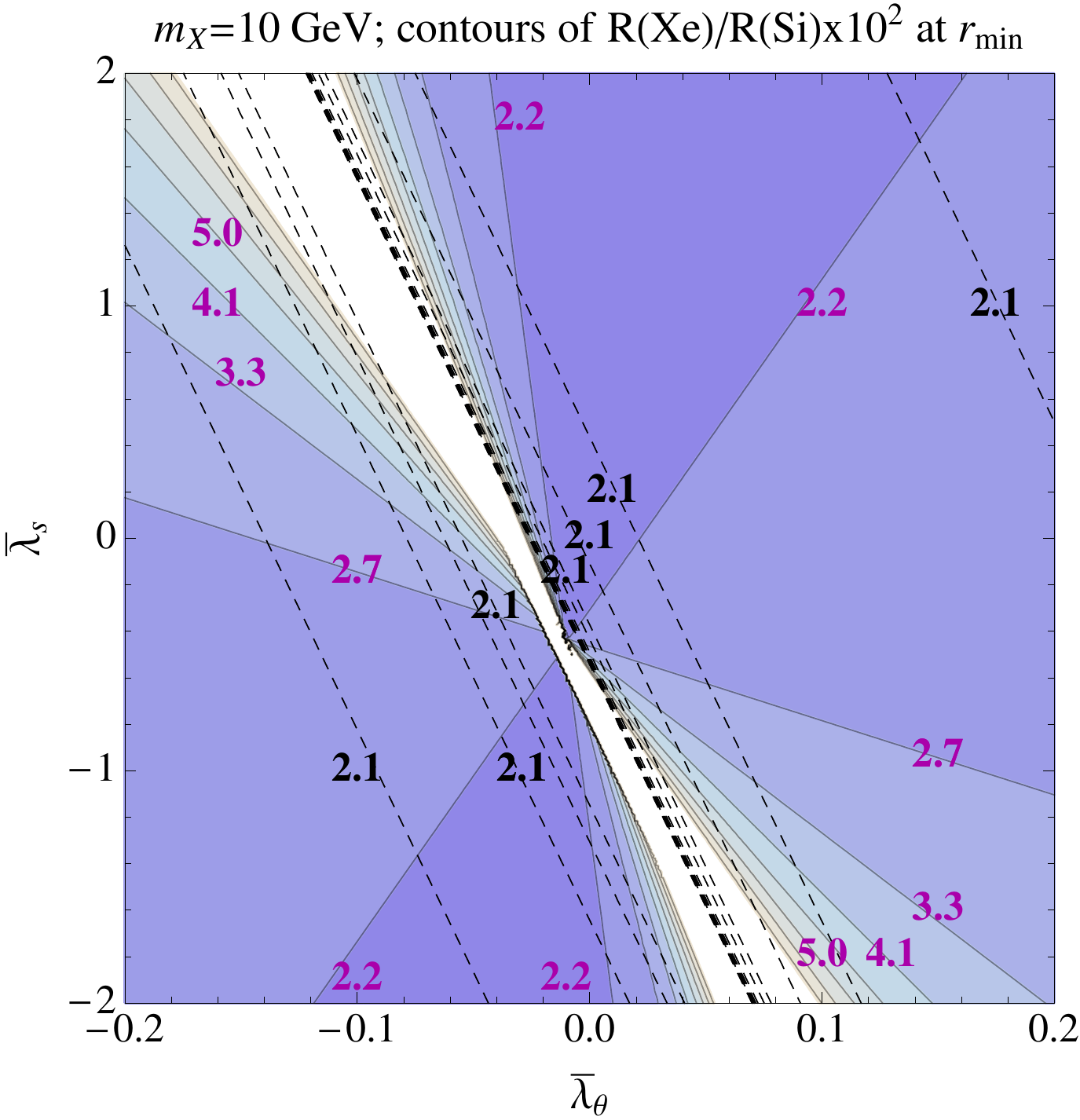}
\vspace{-.3cm}
\caption{\label{fig:pheno3} 
These two panels  show  the behavior of   $\bar{R}^{NLO} (Xe)/ \bar{R}^{NLO} (Si)$  as a function of $\overline{\lambda}_\theta$ and $\overline{\lambda}_s$.  
The left panel shows  contour lines  of constant $r_{\rm min}$, where $r_{\rm min}$ is the location of the minimum of  $\bar{R}^{NLO} (Xe)/ \bar{R}^{NLO} (Si)$ for fixed couplings $\overline{\lambda}_\theta$ and $\overline{\lambda}_s$.
The right panel shows contour lines of constant  $\bar{R}^{NLO} (Xe)/ \bar{R}^{NLO} (Si)$  evaluated at the minimum $r = r_{\rm min}$. 
In both panels the solid lines correspond to full expressions. The dashed lines correspond to the approximate expression of Eq.~(\ref{eq:rate2nlo}) properly averaged over isotopes. In the right panel the red colors give the value of the full expression along that contour, whereas the black colors give the value of the approximate expression. Note that good agreement between the analytic and full expression for $r_{\rm min}$ in the left panel. In contrast, in the right panel the difference in the double ratio between the full and analytic expression becomes $O(1)$ as $r_{\rm min}$ gets large; see Sect. \ref{sect:deg} for more details.  
Central values of the hadronic matrix elements are assumed. 
}
 \vspace{-0.5cm}
\end{center}
\end{figure}

In particular,  the above expressions explain very peculiar degeneracies observed when one scans 
in both the WIMP-quark couplings  $\overline{\lambda}_{\theta,s}$  
and  in the hadronic  and nuclear matrix elements   $\sigma_{\pi N}, t_{u,d}$ (see Figs.~\ref{fig:pheno2},\ref{fig:pheno3}). 
All the  degeneracies  derive from  the relation 
\be
\Delta [ \overline{\lambda}_{s,\theta}, \sigma_{\pi N}, \sigma_{s}, t_{u,d}] ={\rm constant}~.
\label{eq:delta}
\ee
For fixed hadronic matrix elements, this constraint describes a sub-surface in the space of couplings, independent of $r$. Allowing for hadronic uncertainties puffs the surface out into a sub-volume. 
For example, keeping $\lambda_{s} =0$ and $\sigma_{s}, t_{u,d}$ fixed to their central values, 
we obtain very similar results  for the three following choices: 
(1)  $\sigma_{\pi N} = 45$~MeV,  $\overline{\lambda}_\theta = -0.15$ ; 
(2)  $\sigma_{\pi N} = 30$~MeV,  $\overline{\lambda}_\theta = +0.1$ ; 
(3)  $\sigma_{\pi N} = 60$~MeV,  $\overline{\lambda}_\theta = -0.1$. 
They correspond to very close values of $\Delta = 0.147, 0.144, 0.152 $, respectively. 
More generally we show in  Fig.~\ref{fig:pheno3}  the full extent of these degeneracies, using both the full NLO results and the approximate formula Eq.~(\ref{eq:rate2nlo}) (dashed lines), both 
appropriately averaged over isotopes. 
The left panel shows contours of fixed $r_{min}$ (where $\bar{R}^{NLO} (Xe)/ \bar{R}^{NLO} (Si)$ is minimized for fixed couplings $\overline{\lambda}_\theta$ and $\overline{\lambda}_s$).  Here one finds a range of values for $r_{\rm min}$. The right panel shows contours of
$\bar{R}^{NLO} (Xe)/ \bar{R}^{NLO} (Si)$  evaluated at $r_{\rm min}$. Here one finds the double ratio to have only $O(1)$ variation across the plane, demonstrating the existence of other values of $r$, $\lambda_s$ and $\lambda_\theta$ 
having equally good suppression of the relative rate as compared to the canonical IVDM scenario. 
In comparing the two panels note 
the approximate analytic and full numerical expressions have good agreement for contours of $r_{\rm min}$, whereas for the double ratio  $\bar{R}^{NLO} (Xe)/ \bar{R}^{NLO} (Si)[r_{\rm min}]$ there is also good agreement over much of the panel, except in the region where $r_{\rm min}$ becomes large. These two seemingly contrasting features can be easily understood. 
The point is that the numerator of the double ratio is a quadratic form in $r$ and $\Delta$, with slightly different coefficients between the exact and approximate expressions. Since the value of the quadratic form at the minimum is suppressed (with only one isotope it would be zero) through a cancellation between terms that are each large, small differences in the coefficients between the full and approximate expressions lead to larger variation in the value of the minimum, especially as $r_{\rm min}$ becomes large
  ~\footnote{This can be understood in more detail.  Indeed consider a quadratic function $V(r)=ar^2+br+c$. The position of the minimum and value at the minimum are: $r_{\rm{min}}=-b/(2a)$ and $V(r_{\rm{min}})=c-b^2/(4a)$. If we know the coefficients $a,b,c$ only approximately: $a_1=a(1+\epsilon_1), b_1=b(1+\epsilon_2), c_1=c(1+\epsilon_3)$, then the approximate formulas for $r_{\rm{min}}^{\rm{appr}}=r_{\rm{min}}(1+\epsilon_2-\epsilon_1)$ and $V(r_{\rm{min}})^{\rm{appr}}=V(r_{\rm{min}})(1+\epsilon_3)+ar_{\rm{min}}^2(\epsilon_1-2\epsilon_2+\epsilon_3)$ and thus if $a r_{\rm{min}}^2>>V(r_{\rm{min}})$ the value at the minimum cannot be resolved by an approximate formula. In reality in right panel of Fig. 3 the ratio of two quadratic equations is minimized, but the conclusions from our toy model apply. 
  }.

An approximate degeneracy also passes through the canonical IVDM point having $r=-0.7$ and $\overline{\lambda}_s=\overline{\lambda}_\theta=0$. This  point has $\Delta =0$, which selects $\delta f \simeq -0.118$ MeV.  One finds almost perfect degeneracy in the degradation variable along this line, provided $\overline{\lambda}_s < O(1)$. Values of couplings along this line will provide as good a fit to the direct detection data as the canonical point. For larger values of $\overline{\lambda}_s$ the slope terms become important and the degeneracy  weakens. 

This analysis illustrates an important point: hadronic uncertainties affect the extraction of quark-WIMP couplings 
from phenomenologically interesting regions  in the  $\sigma_p-r$ plane. 
In turn, this affects  other aspects of WIMP phenomenology such as indirect detection or collider searches.


Finally,  the above expressions also show how the effect of chiral corrections on the location of the minimum 
is amplified. For example a typical chiral correction  $\Delta \sim 0.15$ implies that for both Ge and Xe  the second factor in Eq.~(\ref{eq:rmin}) is about 1.8 and nearly
the same for both elements because they each have $\bar{Z} \simeq 0.4$.  The amplification arises from the factors of $\bar{Z}, 1- \bar{Z}$ and from the fact that the corrections to numerator and  denominator have the opposite sign.    %

Generalizing the current NLO analysis, one can show that {\it to all orders in the chiral expansion}
the rate (Eq.~\ref{eq:rate1nlo}) takes the form of Eq.~(\ref{eq:rate2nlo}), with the replacement 
$\Delta \cdot  F_{\pi \pi} (E_R)  \to   \Delta_\chi  (E_R; \overline{\lambda}_{s,\theta})$ , 
and $\Delta_\chi$ depending  non-trivially on  $E_R$ 
and $\overline{\lambda}_{s,\theta}$ 
(keeping $O(\xi)$ terms  results in two different functions  $\Delta_{\chi}^{(1),(2)}$ in the two terms of   Eq.~(\ref{eq:rate2nlo})).  
Now, as long as  $\Delta_\chi$ has a well behaved expansion 
(i.e. there are no dynamical enhancements on the nuclear side, which we do not expect for scalar operators), 
then the corrections to the rate and key quantities such as $r_{\rm min}$,  $R(r_{\rm min})$  are well behaved. 
We therefore conclude that our analysis is robust against higher order corrections in the chiral expansion.

\section{CDMS-Si vs XENON and LUX  at NLO}
\label{sect:pheno2}

Throughout, we will assume the Standard Halo Model (SHM), which posits $\rho_{0} = 0.3$ GeV/${\rm cm}^{3}$ and a Maxwell-Boltzmann velocity distribution with variance $v_{0} = 220$ km/s, earth-dark matter relative velocity $v_e=220$ km/s, and escape velocity $v_{esc} = 544$ km/s. In this letter, we will not consider the sizeable uncertainty in the details of the local DM halo. The interested reader can consult previous direct detection studies which have examined in detail the astrophysical uncertainties afflicting direct detection experiments~\cite{Green:2007rb,Green:2008rd,Peter:2009ak,Strigari:2009zb,McCabe:2010zh,Fox:2010bz,Fox:2010bu,Frandsen:2011gi,Kavanagh:2012nr,Friedland:2012fa,Frandsen:2013cna,DelNobile:2013cva,DelNobile:2013gba}.  
We summarize below the key features of our fitting procedure: 

{\bf CDMS Si:}
~We use the acceptance from \cite{Agnese:2013rvf}  and a total exposure of 140.2 kg-days. We consider an energy interval [7,100] keV and bin the data in 2 keV intervals. The 3 candidate events appear in the first 3 bins. Following \cite{Frandsen:2013cna}, we take the normalized background distributions from~\cite{cdmssislides} and rescale them so that neutrons contribute 0.13 events, Pb recoils 0.08 events, and the surface event background 0.41 surface events. To find best-fit regions we obtain the likelihood function and simply plot constant values of the likelihood that would correspond to 68\% and 90\% CL region under the assumption that the likelihood distribution is Gaussian.

{\bf CDMS Ge:} 
~The CDMS collaboration performed a dedicated analysis of their detector at low threshold energy  \cite{Ahmed:2010wy}.
The experiment has a signal region from 2 keV to 100 keV. Following \cite{Farina:2011pw} and  \cite{Frandsen:2011gi}, we set limits using only one of their Ge detectors - T1Z5 - that apparently has the best quality data.   We use the efficiencies and total exposure provided by the
  supplemental information to \cite{Ahmed:2010wy}. The total exposure of this detector was 35 kg--days. 
To account for the finite energy resolution of the detector, the energy of the nuclear recoil is smeared according to 
 \cite{Savage:2008er}
with an energy resolution $\Delta E= 0.2 \sqrt{E/\hbox{keV}} $ keV  \cite{Frandsen:2011gi}. This experiment saw 36 events in their signal region whose origin remains undescribed. To set a conservative upper limit we attribute all of these events to signal - following the experimental collaboration and other theory papers \cite{Farina:2011pw,Frandsen:2011gi}. Using Poisson statistics a 90\% C.L. signal upper limit of 44 events is obtained.

For the Xenon10, Xenon100 and LUX experiments we follow \cite{Aprile:2011hx} and 
convolve the energy-rate $dR/dE$ with a Poisson distribution in the number of photoelectrons or electrons detected. The mean number of electrons expected $\nu(E)$ is specific to each experiment, depending on energy-dependent light or electron yields, and on scintillation efficiencies.
 
 {\bf LUX:} 
~The first data release from LUX \cite{Akerib:2013tjd} has an exposure of 10,065 kg--days. An upper limit of 2.4 signal events for $m_{DM} < 10$ GeV is reported \cite{lux-talk},  with up to 5.3 events allowed for larger masses. We  conservatively apply a limit of 2.4 signal events to the whole mass range $m_{DM} \in (5,30)$ GeV. We use the acceptance provided by \cite{Akerib:2013tjd}. 
We use the energy-dependent light-yield $L_y$ presented in \cite{lux-talk}, including a sharp cutoff at 3 keV. We use the  scintillation efficiency $L_{eff}$ provided by \cite{pc-sorensen}. 
After convolving, we then sum over the S1 signal region (2,30), finding good agreement with the LUX limits \cite{Akerib:2013tjd}. 
Smearing the number of photoelectrons produced with a gaussian to model the response of the detector, as in \cite{Aprile:2011hx}, with a variance of 0.5 PE (photoelectrons), does not appreciably affect our limits. 

\begin{figure}[!t]
\begin{center}
\includegraphics[width=0.3\textwidth]{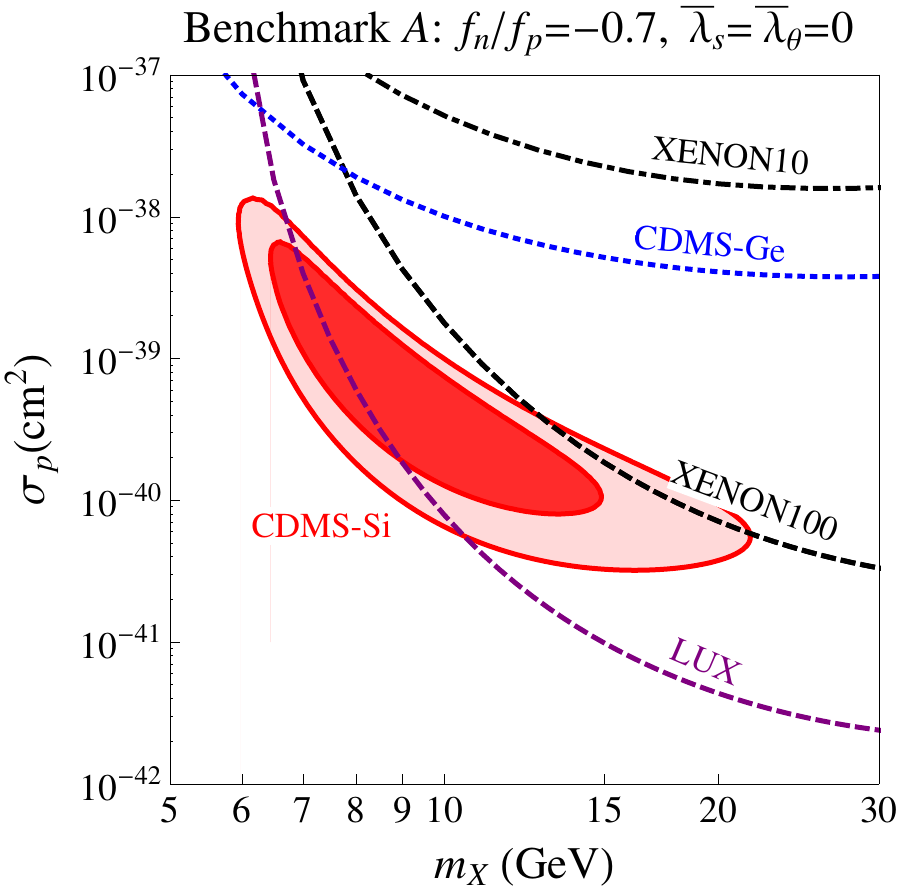}
\includegraphics[width=0.3\textwidth]{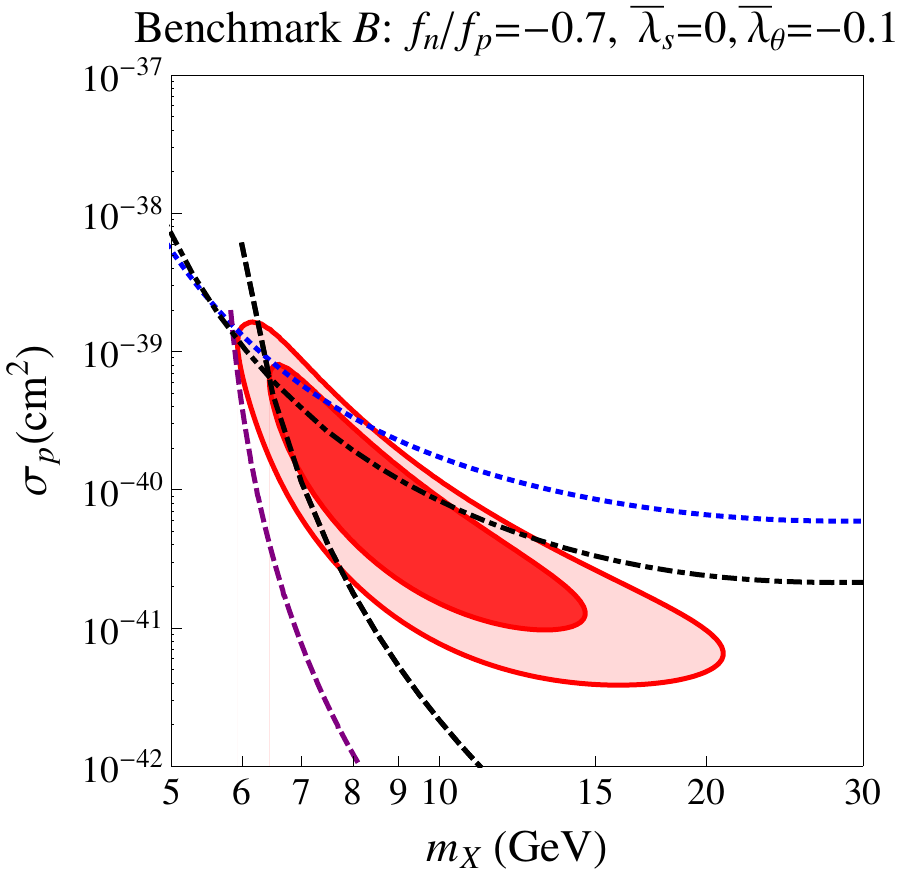}
\includegraphics[width=0.3\textwidth]{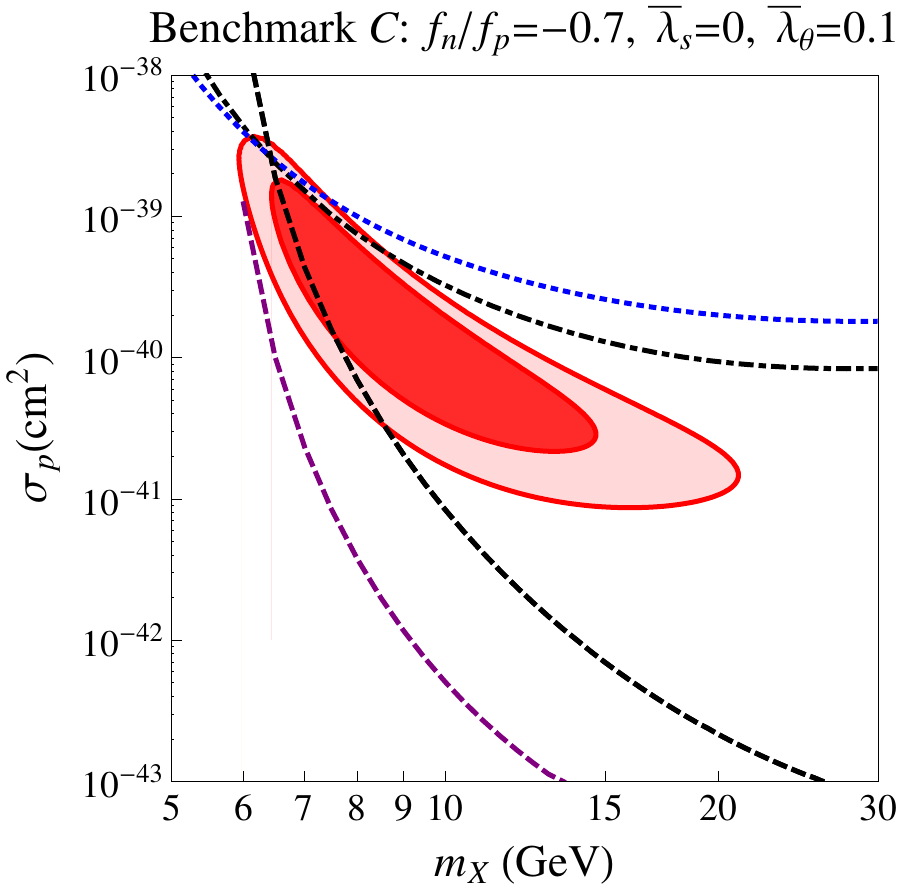}
\caption{\label{fig:pheno4} 
Best-fit CDMS-Si (contours at 68\% and 90\% CL) and XENON/CDMS-Ge/LUX  exclusions (at 90 \% CL) under differing assumptions labelled on the top of each panel. In all cases, we have set $\lambda_{s} = 0$ and used central values of the hadronic matrix elements. The left-hand panel shows the "conventional" IVDM point, reproducing results found in \cite{Gresham:2013mua}. The middle and right panel show the same $r=-0.7$ point with small amounts of $\overline{\lambda}_{\theta}$ turned on. Note that for both points the region allowed in the left panel is now excluded.}
 \vspace{-0.5cm}
\end{center}
\end{figure}

{\bf Xenon10:} 
~While the values of the electron yield $Q_y(E)$ at low energies are controversial, here we simply adopt the collaboration's parameterization from Fig.1 of \cite{Angle:2011th}, 
assuming a sharp cutoff to zero at 1.4 keV nuclear recoil energy. 
Their signal region is from 5 electrons to $\approx$ 35 electrons, corresponding to nuclear recoils of $\approx$1.4 keV to 10 keV, and has an effective exposure of 6.25 kg--days.  A limit is obtained using Poisson statistics with 23 events expected and 23 detected, allowing 9.2 events.
 
{\bf Xenon100:} 
~We use the mean $\nu(E)$ characterized by \cite{Aprile:2011hx}.
For the scintillation efficiency $L_{eff}$ we use the 
efficiency used in Xenon100's  225-live-day analysis \cite{Aprile:2012nq}, that can be found in 
Fig.1 of ref. \cite{Aprile:2011hi} and  includes a linear extrapolation to 0 for $E$ below 3 keV. 
The response of the detector is modeled by a Gaussian smearing  with a mean $n$ and variance $\sqrt{n}  \sigma_{PMT}$ with $\sigma_{PMT}=0.5$ PE \cite{Aprile:2011hx}. The smearing also includes a photoelectron-dependent acceptance, which we parameterize from Fig.1 of \cite{Aprile:2012nq}.  
To get the total rate we then sum the differential rate over the signal region - which for the analysis in  \cite{Aprile:2012nq} corresponds to $S1 \in(3,30)$ PE - and use a total exposure of  $225\times34$ kg-days \cite{Aprile:2012nq}.  
We then use Poisson statistics to obtain a 90$\%$ C.L.  upper limit where 1 background event is expected and 3 observed.   

In general we find our exclusions and best-fit region of LO analysis for $r=1$ -- the only point we can compare to --  have good agreement with those of the experimental collaborations. 
 
Let us now turn to discussing fits to the benchmark points shown in Fig. \ref{fig:pheno1}. In the  three panels of Fig.~\ref{fig:pheno4} we present our NLO results for $r=-0.7$ and $\overline{\lambda}_s=0$, $\overline{\lambda}_{\theta}=0,\pm 0.1$. Our fit for  $r=-0.7$ and  $\overline{\lambda}_s=0$, $\overline{\lambda}_{\theta}=0$ (Benchmark {\it A}) agrees well with the LO fits in the literature (see e.g. \cite{Frandsen:2013cna,Feng:2013vod}, and recently, \cite{Gresham:2013mua}).   The $r=-0.7$ NLO fit with $\overline{\lambda}_{\theta} = \overline{\lambda}_{s} = 0$ is essentially identical to the LO fit, since at this benchmark point the NLO corrections are accidentally small. The 
 smallness of NLO corrections for these coupling values is discussed previously in Sect. \ref{sect:pheno1}.  As one can see from all panels in Fig.~\ref{fig:pheno4} we find that although these benchmarks have the same values of $r$, they lead to qualitatively different fits as expected, with a valid region in the parameter space consistent with CDMS-Si signal and LUX bound only for $\overline{\lambda}_s=0$, $\overline{\lambda}_{\theta}=0$ . Even a relatively small heavy quark coupling, 
 $\lambda_{\theta} = \pm 0.1 \lambda_{u}$, results in a completely excluded region with $r=-0.7$. Thus for $r=-0.7$ to remain a possibility for improving the compatibility between CDMS-Si and the null LUX searches, one must examine models with either (i) strongly suppressed second and third generation couplings, or (ii) those lying on the $\delta f \simeq 0$ degeneracy, as described in Sect. \ref{sect:deg}. 

\begin{figure}[!t]
\begin{center}
\includegraphics[width=0.3\textwidth]{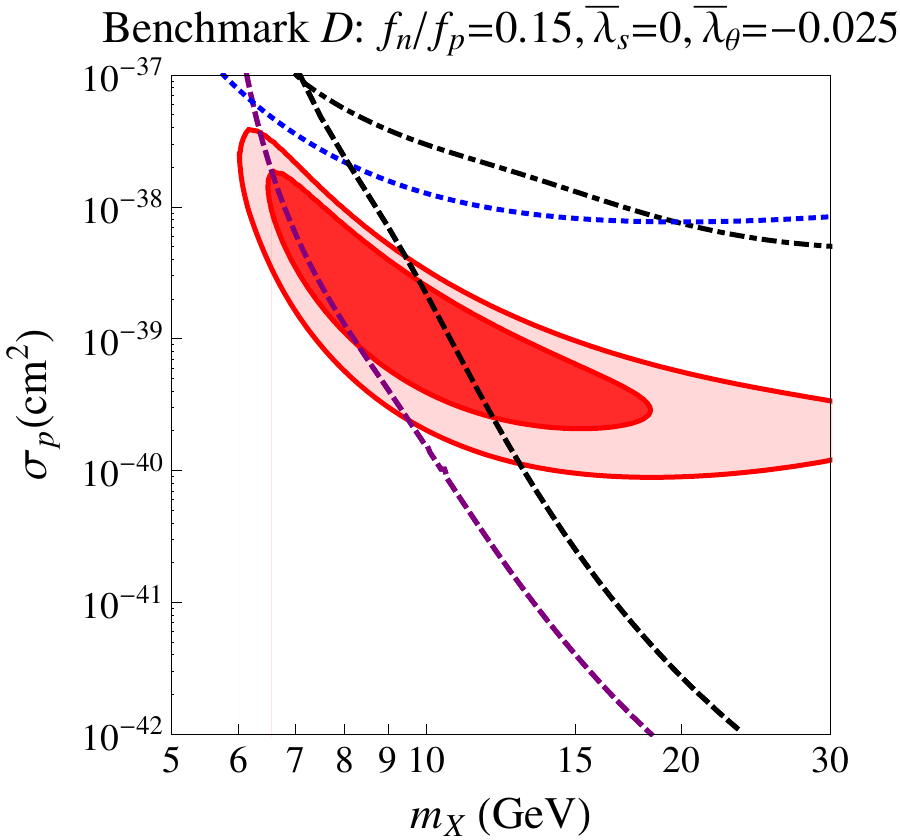}
\includegraphics[width=0.3\textwidth]{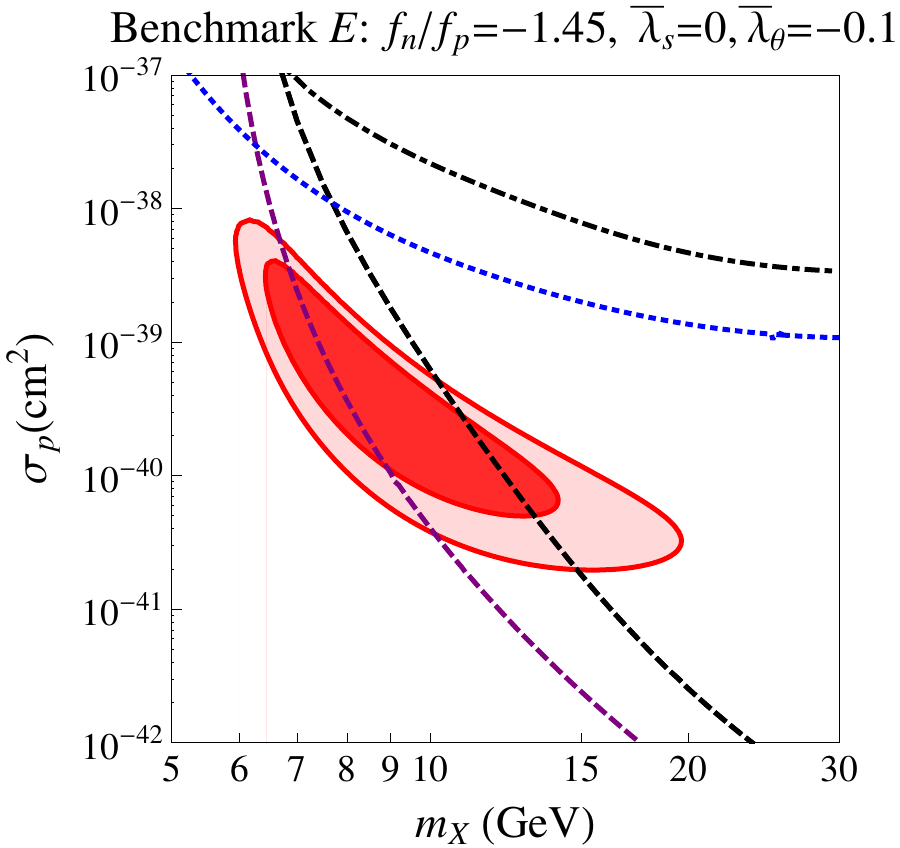}\\
\includegraphics[width=0.3\textwidth]{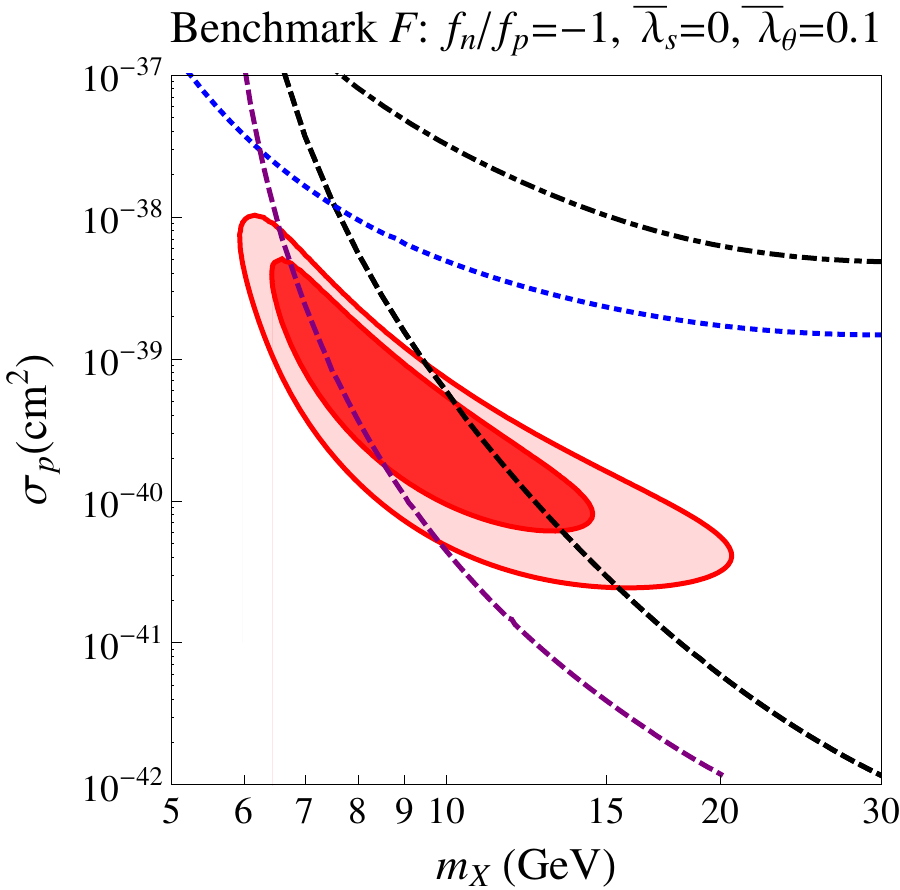}
\includegraphics[width=0.3\textwidth]{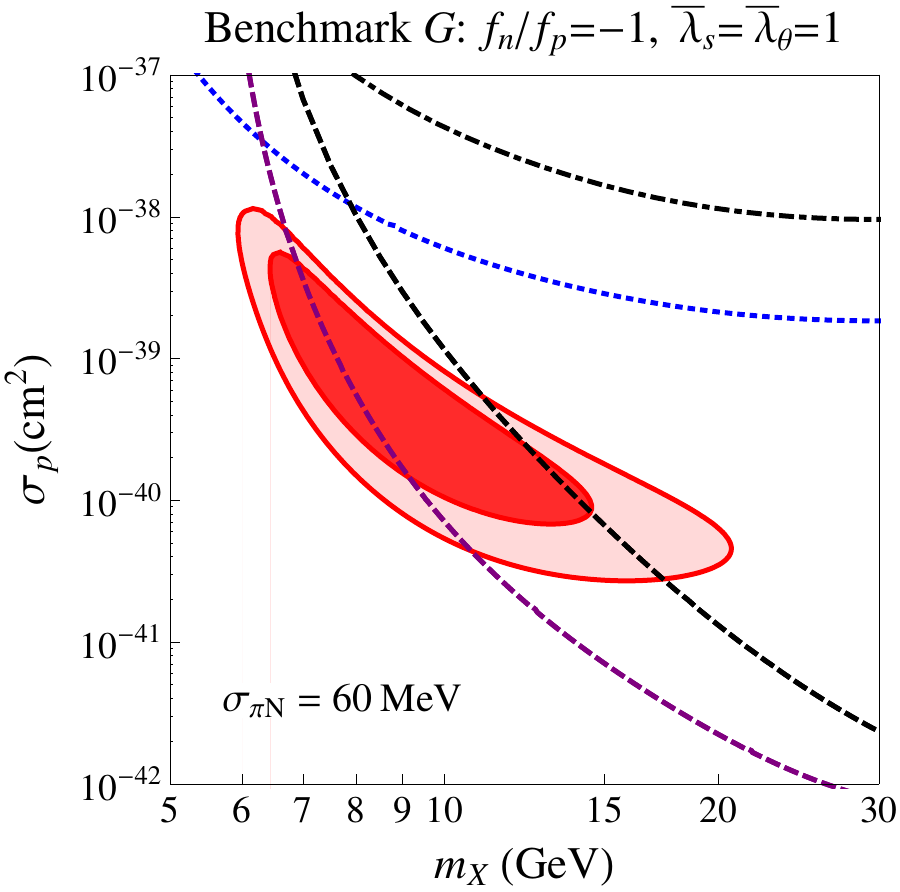}
\caption{\label{fig:pheno5} 
Same assumptions as in Fig. \ref{fig:pheno3} with unconventional choices of $r$ that are excluded at LO.  Note especially the panel on the bottom right-side which compared to the other panels has a different choice of $\sigma_{\pi N}=60$ MeV. The allowed and excluded regions are practically identical to the panel on the bottom left-side having the same value of $r$. The similarity of these two panels illustrates the interplay of allowed or excluded regions and uncertainties in the hadronic parameters.}
 \vspace{-0.5cm}
\end{center}
\end{figure}

Given this tension with the $r=-0.7$ solution at NLO, one may wonder if new solutions with different values of $r$ arise. This indeed seems plausible given the results of Sec.~\ref{sect:pheno1}. Inspecting the left panel of Fig.~\ref{fig:pheno1} we see three choices of parameters that may result in an improved compatibility between LUX and CDMS-Si: (1) Benchmark {\it D}: $\lambda_{\theta} = -0.025 \lambda_{u}$ with $r = +0.15$, (2) Benchmark {\it E}: $\lambda_{\theta} = -0.1 \lambda_{u}$ with $r = -1.45$, and (3) Benchmark {\it F}: $r = -1$ for $\lambda_{\theta} = +0.1 \lambda_{u}$.   This observation motivates the choice of Benchmarks $D$, $E$ and $F$ whose fits are shown in Fig. \ref{fig:pheno5}. We see that these very different choices of $-1.45 \lesssim f_{n}/f_{p} \lesssim .15$ can result in a comparable reduction in tension between the Xenon based experiments and CDMS-Si. In the absence of NLO corrections, these benchmarks would be strongly excluded.  

Lastly, we choose Benchmark {\it G} ($\overline{\lambda}_{\theta} =\overline{\lambda}_{s}=1$ with $r =-1$) to illustrate one of the degeneracies discussed in Sect. \ref{sect:deg}. The fit with this set of parameters is illustrated in the bottom right panel of Fig.~\ref{fig:pheno5}. This final benchmark is chosen with $\sigma_{\pi N} = 60$ MeV, such that it is roughly degenerate with Benchmark {\it F}. Upon inspection of the  fits resulting from the two benchmarks, we see that indeed all the experiments have nearly identical sensitivities.  This final benchmark requires $\sigma_{\pi N}$ to be high in order to remain consistent with the constraints from LUX, and is completely excluded at 90$\%$ CL with $\sigma_{\pi N}$ at its central value of 45 MeV.

\section{Conclusions}
\label{sect:conclusions}
The  CDMS-Si data remain intriguing and may point to a DM candidate with couplings to quarks that are isospin-violating. For a representative case of scalar-mediated DM-quark interactions, we have studied the effect of  long-distance QCD corrections for IVDM models. We use  chiral EFT  and connect the short-distance coefficients directly to the DM-nucleus cross section.

At leading order in chiral power counting it is well-known only two short-distance parameters appear, $r$ and $\sigma_p$. At next-to-leading order, however, for a scalar operator two additional parameters appear. We choose for convenience the following independent parameters $r,\sigma_p,\,\overline{\lambda}_s,\overline{\lambda}_{\theta}$, 
that can all take arbitrary values. 
In the limit of light DM particles,  the  chiral corrections are dominated by the two-nucleon amplitude, for which more work beyond the nuclear shell model 
would be highly desirable.  
We find that for a broad set of values of extra parameters $\overline{\lambda}_s$ and $\overline{\lambda}_{\theta}$ qualitative changes for IVDM phenomenology occur. These can be divided into two categories. 

In the first category, the standard $r=-0.7$ value for IVDM models generically fails to reconcile the LUX exclusion with the CDMS Si best-fit region. This happens because generically the NLO corrections wash out the effect of the LO tuning of Xenon signal. It should be noted however, that for special scenarios, when in the low-energy theory either the DM {\it{only has couplings to the first generation quarks or has couplings lying on 
the $\delta f \approx 0$ degeneracy}} (see Sec.~\ref{sect:deg}), we find that NLO corrections are small, which can be seen from the left panel in Figure~\ref{fig:pheno1}. 
This situation is quite special, as can be seen from the same figure:  
by turning up $\lambda_\theta$ by only $10\%$ of the value for $\lambda_u$, the value for $r$ shifts by a number of the order of 1. 
Consistently, in  Figure~\ref{fig:pheno4} we see that  holding $r=-0.7$ fixed while turning $|\overline{\lambda}_\theta|$ from 0 to $0.1$ results in a complete exclusion. 

In the second category, new values of the parameter $r\ne -0.7$, that are excluded by the leading order analysis, can at NLO partially  reconcile the LUX and CDMS experiments, though strong tension remains. We find that values as low as $r=-1.4$ can be achieved. In fact from  Figures~\ref{fig:pheno1} and \ref{fig:pheno2}  one can see that almost any number $r=-\infty$ to $r=\infty$ is allowed for given (tuned) values of the extra parameters $\overline{\lambda}_s$ and $\overline{\lambda}_{\theta}$.
So there is a manifold of  ``Xenophobic" couplings that extends beyond the canonical point $r = -0.7, \lambda_{s,\theta} =0$ (benchmark $A$). 
This makes the IVDM scenario richer,  but more model-dependent. 
A case in point is provided by the comparison of  benchmarks points $A$ and $G$. 
While leading  to very similar direct detection phenomenology, they have quite distinct short-distance couplings.  
Benchmark $A$, having $r=-0.7$,  corresponds to  $\lambda_d/\lambda_u=\tilde{\lambda}_d/\tilde{\lambda}_u  \simeq -0.5$ and $\lambda_{s,\theta}=0$. 
On the other hand benchmark $G$,  having $r=-1$, has a much larger relative coupling to the heavy quarks or gluons ($\overline{\lambda}_s=\overline{\lambda}_\theta=1$)
and an even  larger relative effective coupling to the down quark: 
$\overline{\lambda}_d \simeq -34$. In terms of ratios of short-distance couplings, $\tilde{\lambda}_s/\tilde{\lambda}_u$=1,
$\lambda_\theta/\tilde{\lambda}_u=0.5$, 
 and $\tilde{\lambda}_d/\tilde{\lambda}_u \simeq -16.5$. 

These results invite a more detailed study of implications of NLO chiral corrections for IVDM. Particularly, since widely different values of the $\lambda_{d}$, $\lambda_s$ and $\lambda_\theta$ couplings can lead to similar direct-detection phenomenology, one expects the constraints and signatures arising from colliders will be important to further distinguish viable scenarios.

\section{Acknowledgements}

We would like to thank Peter Sorensen for invaluable information on the XENON10 and LUX experiments. MG and VC would like to acknowledge  support from the Department of Energy Offices of High Energy  and Nuclear Physics,  and   the Los Alamos LDRD program office. MG would also like to thank the GGI for hospitality and partial support were part of this work was completed. IMS would like to thank Josef Pradler for discussions on the analysis of direct detection data. The CP$^3$-Origins centre is partially funded by the Danish National Research Foundation, grant number DNRF90.

\bibliographystyle{h-physrev}
\bibliography{darkmatter}

\end{document}